\documentclass[aps,superscriptaddress,prc,twocolumn,nofootinbib]{revtex4}

\usepackage{graphicx}
\usepackage{epstopdf}
\usepackage{subfigure}
\usepackage{epsfig}
\usepackage{amsmath,amssymb,amsfonts}
\usepackage{color}

\begin{document}

\title{Heavy quark dynamics and hadronization in ultra-relativistic heavy-ion collisions: collisional vs. radiative energy loss}

\author{Shanshan Cao}
\affiliation{Department of Physics, Duke University, Durham, NC 27708, USA}

\author{Guang-You Qin}
\affiliation{Institute of Particle Physics and Key Laboratory of Quark and Lepton Physics (MOE), Central China Normal University, Wuhan, 430079, China}
\affiliation{Department of Physics and Astronomy, Wayne State University, Detroit 48201, USA}

\author{Steffen A. Bass}
\affiliation{Department of Physics, Duke University, Durham, NC 27708, USA}

\date{\today}

%%%%%%%%%%%%%%%%%%%%%%%%%%%%%%%%%%%%%%%%%%%%%%%%%%%%%%%%%%%%%%%%%%%%%

\begin{abstract}

We study the dynamics of energy loss and flow of heavy quarks produced in ultra-relativistic 
heavy-ion collisions within the framework of a Langevin equation coupled to a (2+1)-dimensional viscous hydrodynamic model that simulates the space-time evolution of the produced hot and dense QCD matter. The classical Langevin approach is improved such that, apart from quasi-elastic scatterings, radiative energy loss is incorporated by treating gluon radiation as an additional force term. The hadronization of emitted heavy quarks is simulated via a hybrid fragmentation plus recombination model. Our calculation shows significant contribution from gluon radiation to heavy quark energy loss at high energies, and we find the recombination mechanism is important for heavy flavor meson production at intermediate energies. We present numerical results for the nuclear modification and elliptic flow of $D$ mesons, which are consistent with measurements at both LHC and RHIC; predictions for $B$ mesons are also provided.

\end{abstract}

\maketitle

%%%%%%%%%%%%%%%%%%%%%%%%%%%%%%%%%%%%%%%%%%%%%%%%%%%%%%%%%%%%%%%%%%%%%

\section{Introduction}

Ultra-relativistic heavy-ion collisions at the Large Hadron Collider (LHC) and the Relativistic Heavy-ion Collider (RHIC) have revealed many interesting and sometimes surprising phenomena. It is now well established that highly excited QCD matter, usually termed the {\it strongly interacting quark-gluon plasma} (sQGP), is created in these energetic nucleus-nucleus collisions \cite{Gyulassy:2004zy}. This hot and dense matter exhibits many remarkable properties, such as the strong collective flow of the final state particles, small viscosity to entropy ratio $\eta/s$, and the opaqueness to high energy jets and partons. Relativistic hydrodynamic simulations have been highly successful in describing the space-time evolution of these produced fireballs \cite{Teaney:2000cw,Huovinen:2001cy,Hirano:2002ds,Nonaka:2006yn,Song:2007fn,Luzum:2008cw,Qiu:2011hf}.

Heavy quarks serve as excellent probes of the QGP fireball as they are primarily produced from early-state hard scatterings and thus have the potential to probe the whole space-time history of the transient matter. Due to their large masses, heavy quarks are expected to be influenced less by the medium compared to light flavor partons and therefore thermalize slower in the dense medium \cite{Moore:2004tg}. Interestingly, experimental observations have revealed significant suppression and anisotropic flow in high-$p_\mathrm{T}$ heavy mesons and heavy flavor decay electrons \cite{Adare:2010de,Tlusty:2012ix,Grelli:2012yv,Caffarri:2012wz} even though they may not equilibrate with the surrounding medium \cite{Cao:2011et}.
In order to understand these phenomena, it is crucial to investigate in detail how heavy quarks evolve and lose energy inside the hot and dense QGP matter and how they hadronize after traversing the medium.

For studying parton evolution and energy loss in dense QCD matter, two important mechanisms are usually considered: medium-induced gluon radiation and quasi-elastic scattering with background medium partons \cite{Wang:1991xy, Braaten:1991we}. For light flavor partons, medium-induced gluon radiation has been shown to be more important than collisional energy loss, e.g., in the suppression of single or triggered hadron production at high transverse momenta \cite{Qin:2007rn, Wicks:2005gt, Renk:2007id, Zakharov:2007pj, Qin:2009bk}.
For heavy quarks, collisional energy loss is usually considered as the dominant mechanism especially at low energies \cite{Moore:2004tg, Mustafa:2004dr}, due to the large masses of heavy quarks which suppress the phase space of gluon radiation. This is known as the ``dead-cone effect" \cite{Dokshitzer:2001zm}.

In the limit of multiple scatterings where the momentum transfer in each interaction is small, the motion of heavy quarks inside a thermalized medium can be treated as Brownian motion and is usually described by the Langevin equation \cite{Svetitsky:1987gq, Moore:2004tg, Akamatsu:2008ge, Gossiaux:2011ea, He:2011qa, Young:2011ug, Alberico:2011zy,Lang:2012cx}. Such a framework has provided a reasonable description of the suppression and elliptic flow of heavy flavor decay products such as the non-photonic electrons measured by RHIC experiments.
However, when extending to higher energy regimes such as those reached by the LHC experiments, heavy quarks become ultra-relativistic as well and thus are expected to behave similarly as light partons.
In this relativistic limit, collisional energy loss alone may no longer be sufficient for simulating the in-medium evolution of heavy quarks, and radiative energy loss corrections may become significant \cite{Abir:2012pu, Cao:2012au}. The incorporation of radiative energy loss into the calculation of heavy quark evolution has been implemented in frameworks such as Boltzmann-based approaches \cite{Gossiaux:2010yx} and parton cascade models \cite{Uphoff:2012gb}, but is still absent in the Langevin approach.

In this work, we study heavy quark evolution and energy loss in a hot and dense QGP medium within the framework of a Langevin equation. In addition to the drag and thermal forces for quasi-elastic scatterings in the Langevin equation, a recoil term is introduced to describe the force exerted on heavy quarks due to gluon radiation. This recoil force is then related to the medium-induced gluon radiation spectrum, which is taken from a higher-twist energy loss calculation \cite{Guo:2000nz,Majumder:2009ge,Zhang:2003wk,Qin:2012fua}.
Within this improved approach, the evolution of heavy quarks inside QGP fireballs is studied, and the significance of the medium-induced gluon radiation is observed, especially for heavy quarks with large transverse momenta.

The hadronization of heavy quarks is simulated with a hybrid fragmentation plus recombination model. In this application, we adopt a ``sudden recombination"  approach for heavy quark coalescence with light quarks from the QGP medium. This approach was first developed for light hadrons formed out of the bulk matter \cite{Dover:1991zn,Fries:2003kq,Greco:2003mm,Chen:2006vc}, and then applied to heavy flavors \cite{Lin:2003jy,Greco:2003vf,Oh:2009zj} and recently to partonic jet hadronization \cite{Han:2012hp}. This coalescence model does not require the thermalization of the recombining partons and it is straightforward to include mesons and baryons simultaneously, and thus it is convenient for the normalization over all possible hadronization channels. Alternative approach, based on the resonance recombination \cite{vanHees:2005wb,vanHees:2007me,He:2011qa}, has also been applied to the study of heavy flavor dynamics.

The paper is organized as follows. In Sec. II, we will present the calculation of heavy quark production from early state hard scatterings. The simulation of heavy quark evolution and energy loss in a dynamic QGP medium will be described in Section III, and the hadronization of heavy quarks via the fragmentation plus recombination mechanism will be discussed in Section IV. In Section V, we will present our calculations of the nuclear modification factor and the elliptic flow of $D$ mesons and compare them with the available experimental data from both LHC and RHIC. Predictions for future measurements of $B$ mesons will also be provided. A summary and outlook will be given in Sec.\ref{sec:summary}.

\section{Heavy flavor initial production}
\label{HQinitialization}

Heavy quarks are mainly produced via hard scatterings at the early stage of relativistic heavy-ion collisions.
The contributions from other processes such as the ``intrinsic heavy quark process" \cite {Vogt:1992ki,Lin:1994xma}, pre-thermal and thermal production \cite{Levai:1994dx,Zhang:2008zzc,Younus:2010sx}, and in-medium jet conversion \cite{Liu:2008bw,Younus:2010sx} have been studied and shown to give small contributions.
In this work, we calculate initial heavy quark distributions using the leading-order perturbative QCD approach \cite{Owens:1986mp} with the incorporation of $gg\rightarrow Q\bar{Q}$ and $q\bar{q}\rightarrow Q\bar{Q}$ processes. For the calculation of partonic cross sections, we utilize CTEQ for the parton distribution functions \cite{Lai:1999wy} and include the nuclear shadowing effect in nucleus-nucleus collisions using the EPS08 parametrization \cite{Eskola:2008ca}.

\begin{figure}[tb]
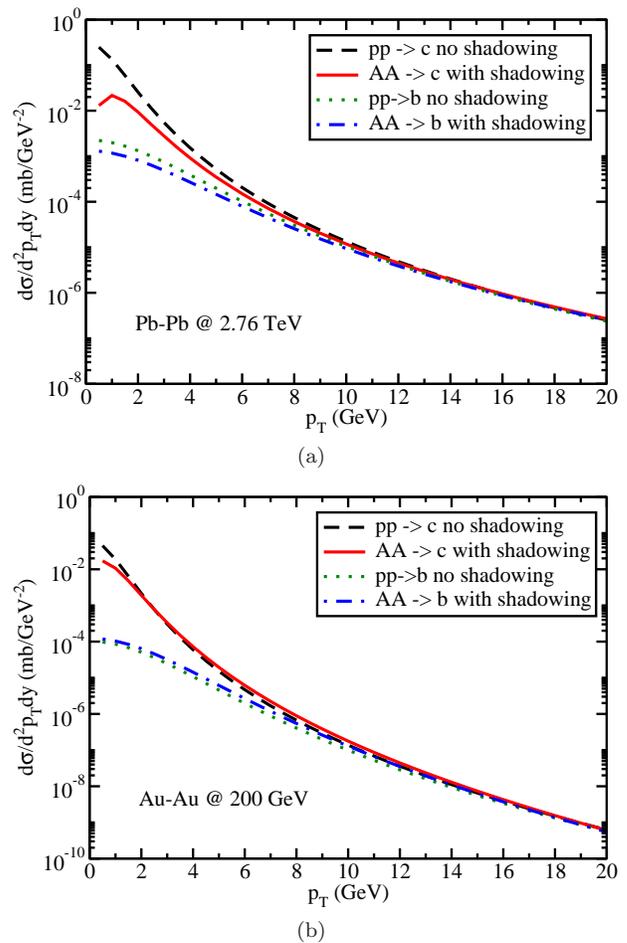

 \subfigure[]{\label{fig:LHCinitial}
      \epsfig{file=LHCppvsAA.eps, width=0.45\textwidth, clip=}}
 \subfigure[]{\label{fig:RHICinitial}
      \epsfig{file=RHICppvsAA.eps, width=0.45\textwidth, clip=}}
 \caption{(Color online) Initial heavy flavor spectra from the leading-order pQCD calculation with and without the nuclear shadowing effect, for the LHC (a) and for RHIC (b).}
 \label{fig:initialspectra}
 \end{figure}

In Fig. \ref{fig:initialspectra}, we show the transverse momentum distributions of initial heavy quarks in proton-proton and binary collision scaled nucleus-nucleus collisions at LHC and RHIC energies.
One can observe from the figure the influence of the nuclear shadowing on the initial heavy quark spectra: it greatly reduces the production rate of charm quark in the low $p_\mathrm{T}$ region; the effect is stronger at the LHC than at RHIC. For the production of the low $p_\mathrm{T}$ bottom quarks, this shadowing effect reduces it at the LHC energy but slightly enhances it at RHIC. These will result in significant effects on the nuclear modification factor $R_\mathrm{AA}$ of heavy flavor hadrons as we will show in later results.

The above calculated distributions are used to sample the transverse momentum $p_\mathrm{T}$ of heavy quarks. The rapidities of initial heavy quarks are taken to be uniformly distributed around the central rapidity region ($-1<\eta<1$). And the spatial distribution of initial heavy quarks is obtained according to the distribution of binary collisions as calculated from the Monte-Carlo Glauber model.

\section{Heavy quark evolution inside a QGP medium}
\label{subsec:HQevolution}

In the limit of small momentum transfer, the multiple scattering of heavy quarks off thermal partons inside a QGP medium can be treated as Brownian motion and thus is typically described using the Langevin equation. In addition to the collisional energy loss resulting from such quasi-elastic scatterings, heavy quarks may also lose energy through  medium-induced gluon radiation. To incorporate both collisional and radiative energy loss experienced by heavy quarks propagating through the dense QGP, we follow our previous studies \cite{Cao:2012jt,Cao:2012au} and modify the classical Langevin equation as follows:
\begin{equation}
\label{eq:modifiedLangevin}
\frac{d\vec{p}}{dt}=-\eta_D(p)\vec{p}+\vec{\xi}+\vec{f}_g.
\end{equation}
The first two terms on the right-hand side are the drag force and the thermal random force from the original Langevin equation, and the third term $\vec{f}_g=-d\vec{p}_g/dt$ is introduced to describe the recoil force exerted on heavy quarks due to gluon radiation, where $\vec{p}_g$ denotes the momentum of radiated gluons.

Assuming the noise term $\vec{\xi}$ is independent of the momentum of each particle, the random force satisfies the following correlation relation:
\begin{equation}
\label{eq:noise}
\langle\xi^i(t)\xi^j(t')\rangle=\kappa\delta^{ij}\delta(t-t'),
\end{equation}
in which $\kappa$ represents the momentum space diffusion coefficient of heavy quarks. 
To simulate the evolution of heavy quarks in medium, we discretize Eqs.(\ref{eq:modifiedLangevin}) and (\ref{eq:noise}) as follows:
\begin{align}
\label{eq:discretizedLangevin}
\vec{p}(t+&\Delta t)=\vec{p}(t)-\eta_D(p)\vec{p}\Delta t+\vec{\xi}\Delta t - \Delta \vec{p}_\mathrm{g}, \\
\label{eq:discretizednoise}
&\langle\xi^i(t)\xi^j(t-n\Delta t)\rangle=\frac{\kappa}{\Delta t}\delta^{ij}\delta^{0n},
\end{align}
where $\Delta \vec{p}_\mathrm{g}$ is the momentum of gluons radiated during the time interval $\Delta t$.

From Eq.(\ref{eq:discretizednoise}), each spatial component of the thermal force $\vec{\xi}$ during a $\Delta t$ can be independently sampled with a Gaussian distribution whose width is $\sqrt{\kappa/\Delta t}$. Meanwhile, we determine the probability of gluon radiation during each $\Delta t$ according to the average number of gluons in this time interval,
\begin{equation}
\label{eq:gluonnumber}
P_\mathrm{rad}(t,\Delta t)=\langle N_\mathrm{g}(t,\Delta t)\rangle = \Delta t \int dxdk_\perp^2 \frac{dN_\mathrm{g}}{dx dk_\perp^2 dt}.
\end{equation}
We choose sufficiently small time steps $\Delta t$ to ensure that the average radiated gluon number is smaller than 1 in a time step $\Delta t$.
In this work, we utilize the results of the higher-twist calculation for the medium-induced gluon spectra \cite{Guo:2000nz,Majumder:2009ge,Zhang:2003wk}:
\begin{eqnarray}
\label{eq:gluondistribution}
\frac{dN_\mathrm{g}}{dx dk_\perp^2 dt}=\frac{2\alpha_s  P(x)\hat{q} }{\pi k_\perp^4} {\sin}^2\left(\frac{t-t_i}{2\tau_f}\right)\left(\frac{k_\perp^2}{k_\perp^2+x^2 M^2}\right)^4,
\end{eqnarray}
where $k_\perp$ is the transverse momentum of the radiated gluon, and $x$ is the ratio between the gluon energy and the heavy quark energy. In addition, $\alpha_s$ is the strong coupling constant, $P(x)$ is the splitting function of the gluon and $\hat{q}$ is the gluon transport coefficient. The gluon formation time $\tau_f$ is defined as $\tau_f={2Ex(1-x)}/{(k_\perp^2+x^2M^2)}$,
with $E$ and $M$ being the energy and mass of heavy quarks.
Note that the quartic term at the end of Eq.(\ref{eq:gluondistribution}) characterizes the ``dead-cone" effect, i.e., the suppression of gluon radiation due to the finite masses of heavy quarks.

At a given time step, Eq.(\ref{eq:gluonnumber}) is used to determine the probability of radiating a gluon. If a gluon is formed, its energy and momentum will be generated with the Monte-Carlo method according to the gluon radiation spectrum in Eq.(\ref{eq:gluondistribution}). After a gluon is radiated from the heavy quark, the initial time $t_i$ in the equation is reset to zero so that the probability of radiating the next gluon starts to accumulate again with time. Note that the framework we describe here does not necessarily require the higher-twist formalism -- other energy loss formalisms can be used as well, as long as they provide the distributions for both energy and transverse momentum of the radiated gluons.

In the literature, the spatial diffusion coefficient $D$ is usually quoted for heavy quark calculations, and related to the momentum space diffusion coefficients in the above equations via
\begin{equation}
D\equiv\frac{T}{M\eta_D(0)}=\frac{2T^2}{\kappa}.
\end{equation}
Meanwhile, we have the relation $\hat{q}=2\kappa C_A/C_F$, where $C_F$ and $C_A$ are color factors for quarks and gluons.
With such a setup, we only have one free parameter in the Langevin framework as described above.
Throughout our calculation presented in this work, the spatial diffusion coefficient of heavy quark is set as $D=6/(2\pi T)$, which corresponds to a gluon transport coefficient $\hat{q}$ around 3~GeV$^2$/fm at a temperature of $T=400$~MeV.

For the classical Langevin equation without the contribution from gluon radiation, we have the fluctuation-dissipation relation:
\begin{equation}
\eta_D(p)=\frac{\kappa}{2TE}.
\end{equation}
For the simulation of radiative energy loss due to medium-induced gluon radiation, we impose a lower cut-off $\omega_0=\pi T$ for the gluon energy to take into account the balance between gluon emission and absorption processes. Below such a cut-off, the gluon emission is disabled and the evolution of heavy quarks with low energies is completely controlled by quasi-elastic multiple scattering. Such a treatment for medium-induced gluon radiation ensures that heavy quarks achieve thermal equilibrium after sufficiently long evolution times.

\begin{figure}[tb]
  \epsfig{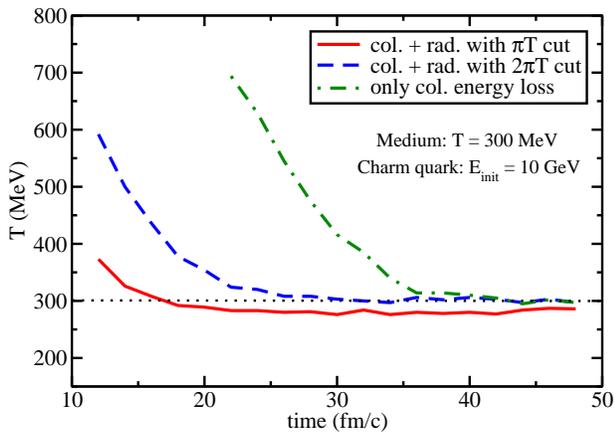}
  \caption{(Color online) Thermalization process of charm quarks in a static medium.}
 \label{fig:thermalization}
\end{figure}

In Fig. \ref{fig:thermalization} we provide a numerical check of the thermalization process of charm quarks according to the modified Langevin equation. The charm quarks are all initialized with an energy of 10~GeV and then evolve inside an infinite and static medium with a constant temperature of 300~MeV. The temperature parameter of the charm quark ensemble is extracted from their energy spectrum utilizing the method developed in our previous work \cite{Cao:2011et}. As is shown, if there is only collisional energy loss, the temperature parameter of the charm quarks evolves to the medium temperature as expected. We also examine such thermalization behavior when the gluon radiation is introduced. If the energy cutoff for the gluon radiation is large enough, e.g., $2\pi T$ in the plot, the heavy quarks will eventually equilibrate with the medium temperature. For the choice of $\pi T$, an equilibrium can still be achieved, with the only difference that the equilibrium temperature is shifted by a small amount, approximately 20~MeV below the medium temperature.

In this work, the lower cut for radiative gluon energy is taken to be $\pi T$, which is the typical energy of the gluons in the thermalized QGP medium. Such choice introduces 5-10\% uncertainty in the equilibrium temperature, but should not substantially influence the description of heavy flavor observables presented here. Additionally, if one considers all sources of experimental and theoretical uncertainties, such as those in hydrodynamic initial conditions and the nuclear shadowing effect, it might not be necessary to artificially increase the energy cut for gluon radiation merely for the exact preservation of the original detailed balance. A more rigorous treatment would incorporate the absorption process as well in the above simulation of gluon radiation. Such an effort has already been explored in the context of light parton radiative energy loss \cite{Wang:2001cs} and will be pursued in a future study.

With the modified Langevin framework described above, we may simulate the heavy quark evolution inside the hot and dense QCD matter created in relativistic heavy-ion collisions. In this work, the space-time evolution profiles of the QGP fireballs at LHC and RHIC are generated with a (2+1)-dimensional viscous hydrodynamic model, which was developed by Song \cite{Song:2007fn,Song:2007ux} and has recently been modified by Qiu and Shen for increased numerical stability \cite{Qiu:2011hf}. Here we employ the code version and parameter tunings that were previously used in Ref. \cite{Qiu:2011hf}. In the following calculation, the MC-Glauber initialization is adopted for the hydrodynamic calculation if not otherwise emphasized. The hydrodynamic model provides the space-time evolution of the temperature and collective flow profiles of the thermalized medium. For every Langevin time step, we boost the heavy quark into the local rest frame of the fluid cell through which it travels. In the rest frame of the fluid cell, the energy and momentum of the heavy quark is updated according to the aforementioned method before it is boosted back to the global center-of-mass frame where it streams freely until the next time step.

In the simulation of heavy quark evolution, we assume they stream freely prior to the initial time $\tau_0=0.6$~fm/c, at which the hydrodynamic evolution commences. We neglect the possible energy loss in the pre-equilibrium stage, which is expected to give a small contribution in this short period of time compared to the much longer history of the fireball evolution. Once they enter a fluid cell with a local temperature below $T_\mathrm{c}$ (165~MeV here), heavy quarks hadronize into heavy flavor hadrons.

\begin{figure}[tb]
 \subfigure[]{\label{fig:cEloss}
      \epsfig{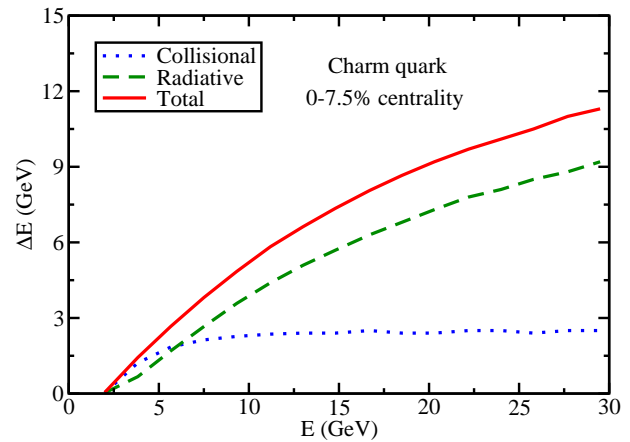}}
 \subfigure[]{\label{fig:bEloss}
      \epsfig{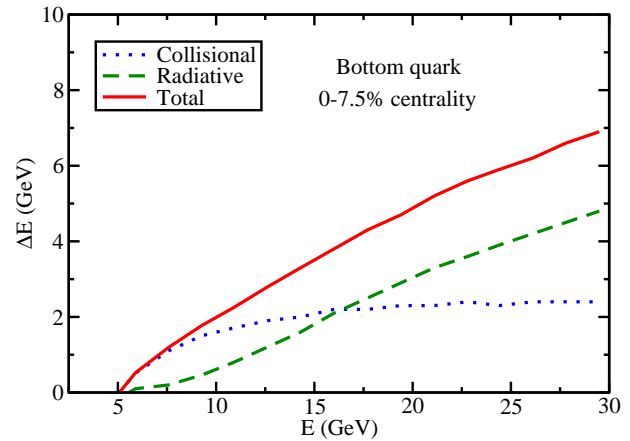}}
 \caption{(Color online) Comparison of radiative and collisional energy losses for charm (a) and for bottom (b) quarks.}
 \label{fig:eloss}
\end{figure}

In Fig. \ref{fig:eloss}, we compare the energy loss of heavy quarks between two different mechanisms after they traverse a QGP medium created in the central Pb-Pb collisions at the LHC. The x-axis represents the initial energy of heavy quarks, and the y-axis denotes the total energy loss.
As is shown, quasi-elastic scatterings dominate the heavy quark energy loss at low energy regime, while medium-induced gluon radiation dominates at high energies. The crossing points are around 6~GeV for charm quarks and 16~GeV for bottom quarks. These results indicate that collisional energy loss alone may provide good descriptions for the heavy flavor measurements at RHIC energies, but will become insufficient when we extend to higher energies, such as those reached by the LHC.

\section{Heavy flavor hadronization}
\label{subsec:HQhadr}

There are two typical hadronization mechanisms for heavy quarks to produce heavy flavor hadrons: high momentum heavy quarks tend to fragment directly into hadrons, while it is more probable for low momentum quarks to hadronize through coalescence (or recombination) with light partons from the thermalized medium. In this work, we combine these two mechanisms for heavy quark hadronization after they travel outside the QGP matter. The fragmentation processes can be simulated by Pythia 6.4 \cite{Sjostrand:2006za} with its ``independent fragmentation model". In the following, we provide details of our implementation of the coalescence processes following the ``sudden recombination" model developed in Ref. \cite{Oh:2009zj}.

In the ``sudden recombination'' model, the momentum distributions of produced mesons and baryons are determined by the following expressions,
%\begin{widetext}
\begin{eqnarray}
\label{eq:recombMeson}
\frac{dN_M}{d^3p_M} \!\!&=&\!\!\! \int d^3p_1 d^3p_2 \frac{dN_1}{d^3p_1} \frac{dN_2}{d^3p_2}f^W_M(\vec{p}_1,\vec{p}_2)\delta(\vec{p}_M-\vec{p}_1-\vec{p}_2) \nonumber \\
\frac{dN_B}{d^3p_B} \!\!&=&\!\!\! \int d^3p_1 d^3p_2 d^3p_3 \frac{dN_1}{d^3p_1} \frac{dN_2}{d^3p_2} \frac{dN_3}{d^3p_3}f^W_B(\vec{p}_1,\vec{p}_2,\vec{p}_3) \nonumber \\  && \times \delta(\vec{p}_M-\vec{p}_1-\vec{p}_2-\vec{p}_3).
\end{eqnarray}
%\end{widetext}
$dN_i/d^3p_i$ represents the momentum distribution of the $i$-th valence parton in the recombined meson or baryon. The distribution of heavy quarks can be directly obtained after they travel through the thermalized medium. For light quarks and anti-quarks from the QGP medium, we take the Fermi-Dirac distribution in their local cell frame:
\begin{equation}
\label{eq:FermiDirac}
\frac{dN_q}{d^3p}=\frac{g_q V}{e^{\sqrt{p^2+m^2}/T_\mathrm{c}}+1},
\end{equation}
where a uniform distribution in the position space is assumed inside a volume $V$ and $g_q=6$ denotes the degrees of freedom for each quark flavor. $f^W$ is the Wigner function of the constructed mesons or baryons, denoting the probability for particles to recombine. For a two-particle system, the Wigner function is defined as:
\begin{equation}
\label{eq:WignerMeson}
f_M^W(\vec{r},\vec{q})\equiv N g_M \int d^3 r' e^{-i\vec{q}\cdot\vec{r}'}\phi_M(\vec{r}+\frac{\vec{r}'}{2})\phi^*_M(\vec{r}-\frac{\vec{r}'}{2}),
\end{equation}
where $N$ is a normalization factor and $g_M$ is a statistic factor taking into account the color and spin degrees of freedom.
For instance, $g_M$ is $1/(2\times 3 \times 2\times 3) = 1/36$ for the $D$ meson ground state, and $3/(2\times 3 \times 2\times 3) = 1/12$ for a $D$ excited state. Here $\vec{r}$ and $\vec{q}$ are the relative position and momentum of the two particles in the center of mass frame of the meson:
\begin{align}
\label{eq:transformation}
\vec{r} \equiv & \vec{r}_1'-\vec{r}_2', \nonumber \\
\vec{q} \equiv \frac{1}{E_1'+E_2'} & (E_2' \vec{p}_1'-E_1'\vec{p}_2').
\end{align}
Note that the variables on the right-hand side of Eq.(\ref{eq:transformation}) are defined in the center of mass frame of the two-particle system, i.e., the meson frame. In addition, $\phi_M$ in Eq.(\ref{eq:WignerMeson}) is the meson wave-function, which is approximated by the ground state wave-function of a quantum mechanic simple harmonic oscillator:
\begin{equation}
\phi_M(\vec{r})=\left(\frac{1}{\pi \sigma^2}\right)^{3/4}e^{-r^2/(2\sigma^2)},
\end{equation}
where the width parameter $\sigma$ is related to the oscillator frequency $\omega$ by $\sigma\equiv 1/\sqrt{\mu\omega}$, in which $\mu\equiv m_1m_2/(m_1+m_2)$ is the reduced mass. After averaging over the position space of Eq.(\ref{eq:WignerMeson}), we obtain the following momentum space Wigner function:
\begin{equation}
\label{eq:momentumWigner}
f^W_M(q^2) = N g_M \frac{(2\sqrt{\pi}\sigma)^3}{V} e^{-q^2 \sigma^2}.
\end{equation}
The Wigner function can be generalized to a three-particle system for baryon production by recombining two particles first and then using their center of mass to recombine with the third one. This yields:
\begin{equation}
\label{eq:baryonWigner}
f^W_B(q_1^2,q_2^2) = N g_B \frac{(2\sqrt{\pi})^6(\sigma_1\sigma_2)^3}{V^2} e^{-q_1^2 \sigma_1^2-q_2^2\sigma_2^2},
\end{equation}
where the relative momenta are defined in the center of mass frame of the produced baryon as
\begin{align}
& \vec{q}_1\equiv\frac{1}{E_1'+E_2'}(E_2'\vec{p}_1'-E_2'\vec{p}_2'), \nonumber \\
& \vec{q}_2\equiv\frac{1}{E_1'+E_2'+E_3'}\left[E_3'(\vec{p}_1'+\vec{p}_2')-(E_1'+E_2')\vec{p}_3'\right];
\end{align}
and the width parameters $\sigma_i$'s are given by $\sigma_i=1/\sqrt{\mu_i \omega}$ with
\begin{equation}
\mu_1=\frac{m_1m_2}{m_1+m_2},
\hspace{15pt}
\mu_2=\frac{(m_1+m_2)m_3}{m_1+m_2+m_3}.
\end{equation}
In general, the frequency $\omega$ can be calculated with the charge radius and is different for each hadron. Here, we adopt the average values of 0.106~GeV for charm hadrons and 0.059~GeV for bottom hadrons as tuned in Ref. \cite{Oh:2009zj}. We use a thermal mass of 300~MeV for $u$ and $d$ quarks and 475~MeV for $s$ quarks. Note that heavy quarks are not required to be thermal in this recombination model and their masses are taken as 1.27~GeV for $c$ and 4.19~GeV for $b$ quarks.

We use these Wigner functions Eqs.(\ref{eq:momentumWigner}) and (\ref{eq:baryonWigner}) to calculate the probability for a heavy quark after its medium evolution to produce a hadron through coalescence with the light quarks from the QGP medium at $T_\mathrm{c}$. The overall normalization factor $N$ is determined by requiring the total recombination probability to be 1 for a zero-momentum heavy quark to all possible heavy flavor meson and baryon channels (we include both ground states and first excited states of $D$/$B$, $\Lambda_Q$, $\Sigma_Q$, $\Xi_Q$ and $\Omega_Q$). The value of the normalization factor is obtained using a static medium with an effective temperature of $T_\mathrm{eff}=175~\mathrm{MeV}$. This effective temperature is chosen to take into account the effect of radial flow (around 0.6$c$ at $T_\mathrm{c}$) developed in the hydrodynamic model, and obtained according to the following equation,
\begin{equation}
\label{eq:effectiveT}
\sum_\mathrm{flavors}\int d^3p \frac{g_q V}{e^{E/T_\mathrm{eff}}+1}=\sum_\mathrm{flavors}\int d^3p \frac{g_q V}{e^{p\cdot u/T_\mathrm{c}}+1}.
\end{equation}
With the choice of $T_\mathrm{eff}=175~\mathrm{MeV}$ and $T_\mathrm{c}=165~\mathrm{MeV}$, both sides of Eq.(\ref{eq:effectiveT}) lead to the same parton density: a number density around $0.24/\mathrm{fm}^3$ for $u$ and $d$, and  $0.13/\mathrm{fm}^3$ for $s$ quark.
More discussions about this effective temperature can be found in Refs. \cite{Oh:2009zj,Oh:2007vf}.

\begin{figure}[tb]
  \epsfig{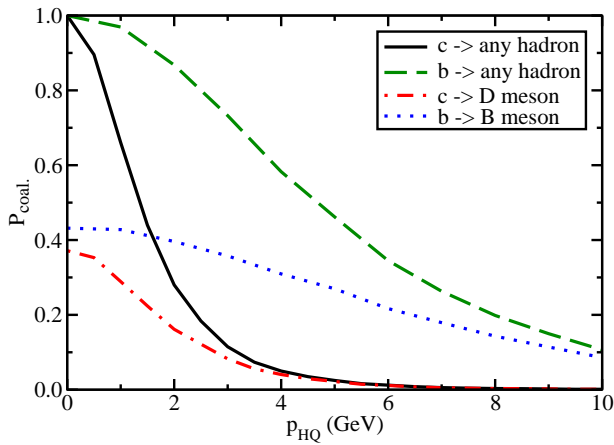}
  \caption{(Color online) The coalescence probabilities for heavy-light quarks as a function of heavy quark momentum.}
 \label{fig:Pcoal}
\end{figure}

Now we may calculate the coalescence probability as a function of heavy quark momentum as shown in Fig. \ref{fig:Pcoal}. 
The recombination probabilities for a charm or bottom quark to all heavy flavor hadron channels and to only $D$ or $B$ meson are shown for comparison.
One observes that for the same $p_\mathrm{T}$, bottom quarks have larger recombination probability than charm quarks to produce heavy flavor hadrons due to their larger masses.
These curves in the plot divide the hadronization of a charm or bottom quark after its medium evolution into three possibilities: recombination to $D$ or $B$ meson, recombination to other hadron channels and fragmentation.
For a charm or bottom quark that is selected for recombination into a $D$ or $B$ meson, a light quark or anti-quark is first generated according to Eq.(\ref{eq:FermiDirac}) in the cell frame of the medium, then boosted to the lab frame to recombine with the charm or bottom quark according to the probability governed by Eq.(\ref{eq:momentumWigner}). If they do not recombine, another light parton will be generated until a meson is formed.
%A charm/bottom quark being chosen to recombine to other hadrons than $D/B$ meson will be not analyzed in this work.
The fragmentation of heavy quarks is achieved via Pythia simulation where the relative ratios between different channels have been properly calculated and normalized.

\begin{figure}[tb]
 \subfigure[]{\label{fig:rec-frag-D}
      \epsfig{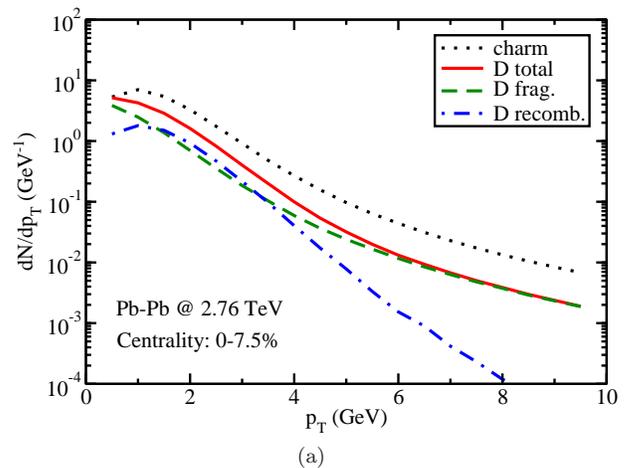}}
 \subfigure[]{\label{fig:rec-frag-B}
      \epsfig{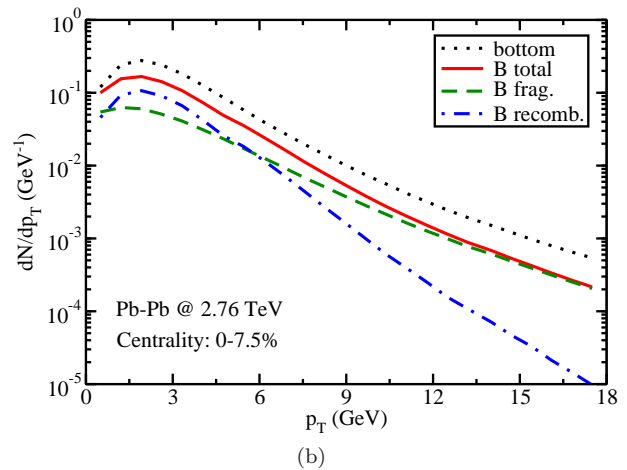}}
  \caption{(Color online) The relative contributions from different hadronization mechanisms to (a) $D$ and (b) $B$ meson production from heavy quarks (normalized to one heavy quark).}
  \label{fig:rec-frag}
\end{figure}

Figure \ref{fig:rec-frag} illustrates the relative contributions from recombination and fragmentation mechanisms to the production of heavy flavor mesons from charm and bottom quarks after they have passed through the thermalized medium. One can see that while the fragmentation dominates the $D$/$B$ meson production at high $p_\mathrm{T}$, the inclusion of the recombination mechanism greatly increases their yield at intermediate $p_\mathrm{T}$. As the recombination mechanism adds a thermal parton to a heavy quark, the momentum distribution of $D$/$B$ mesons through recombination is shifted to the right (higher momenta) compared to charm/bottom quark distribution. Consequently, its contribution to $D$/$B$ meson production at low $p_\mathrm{T}$ is not as significant as at intermediate $p_\mathrm{T}$. Furthermore, due to the larger mass of $b$-quarks, the contribution from the recombination mechanism to $B$ meson production is more prominent than to $D$ meson over a wider $p_\mathrm{T}$ range.

\section{Nuclear modification and elliptic flow of heavy flavor mesons}
\label{sec:results}

After obtaining the final heavy flavor mesons as described in above sections, we calculate the nuclear modification factor $R_\mathrm{AA}$ and the elliptic flow coefficient $v_2$ of heavy flavor mesons which are defined as:
\begin{align}
& R_\mathrm{AA}(p_\mathrm{T})\equiv\frac{1}{N_\mathrm{coll}}\frac{{dN^\mathrm{AA}}/{dp_\mathrm{T}}}{{dN^\mathrm{pp}}/{dp_\mathrm{T}}}, \\
& v_2(p_\mathrm{T})\equiv\langle \cos(2\phi)\rangle=\left\langle\frac{p_x^2-p_y^2}{p_x^2+p_y^2}\right\rangle.
\end{align}
Below we will present the results for both LHC and RHIC and compare with the available experimental data.

\begin{figure}[tb]
  \epsfig{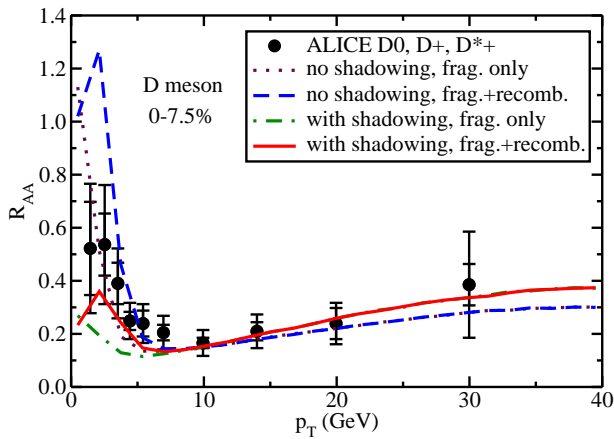}
  \caption{(Color online) Nuclear modification factor $R_\mathrm{AA}$ for $D$ mesons in the most central Pb-Pb collisions at the LHC energies. Different initial production and hadronization mechanisms are compared.}
 \label{fig:LHCRAA}
\end{figure}

In Fig. \ref{fig:LHCRAA} we show the calculation of the $D$ meson nuclear modification factor for the most central Pb-Pb collisions at the LHC.
The impact of nuclear shadowing and recombination on the nuclear modification of $D$ mesons can be clearly seen in the result. 
With the inclusion of the shadowing effect, we obtain a factor of four decrease in the $D$ meson $R_\mathrm{AA}$ at low $p_\mathrm{T}$, while a mild increase is observed at high $p_\mathrm{T}$.
This is due to the fact that the charm quark production is significantly suppressed at low $p_\mathrm{T}$ and slightly enhanced at high $p_\mathrm{T}$ in Pb-Pb collisions relative to binary collision scaled proton-proton collisions, as shown in Fig. \ref{fig:LHCinitial}.
We also observe that fragmentation alone is sufficient to describe heavy quark hadronization above 8~GeV, but in the low and intermediate $p_\mathrm{T}$ region, the recombination of light and heavy quarks becomes important.
This is because the coalescence mechanism brings low $p_\mathrm{T}$ heavy quarks into medium $p_\mathrm{T}$ hadrons by combining a thermal parton from the QGP medium, and therefore decreases the $D$ meson $R_\mathrm{AA}$ near zero $p_\mathrm{T}$ but significantly increases it in the intermediate regime (2-5~GeV).
We see that after the incorporation of the shadowing effect in the initial heavy quark production, the radiative and collisional energy loss mechanisms, and the fragmentation plus recombination mechanism for hadronization, a good description of the $D$ meson $R_\mathrm{AA}$ data measured by the ALICE collaboration is obtained.

\begin{figure}[tb]
 \subfigure[]{\label{fig:LHCv2hadr}
      \epsfig{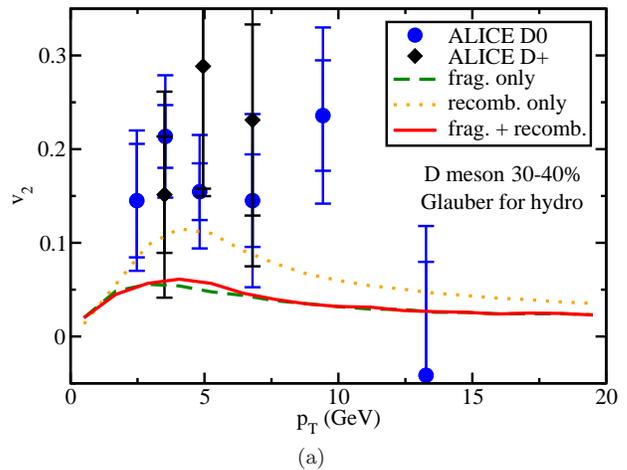}}
 \subfigure[]{\label{fig:LHCv2init}
      \epsfig{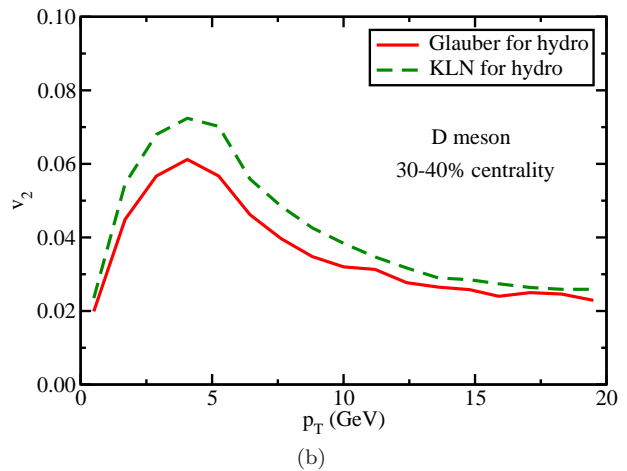}}
 \caption{(Color online) Elliptic flow $v_2$ of $D$ mesons at the LHC.  Different initial conditions and hadronization mechanisms are compared.}
 \label{fig:LHCv2}
\end{figure}

In Fig. \ref{fig:LHCv2}, we show our calculation of the $D$ meson elliptic flow $v_2$. The result for different hadronization scenarios are presented for comparison in Fig. \ref{fig:LHCv2hadr}. For the pure fragmentation process, we set the Wigner function $f^W$ to be 0 to shield all hadronization channels through coalescence, while $f^W$ is taken as 1 for the pure recombination process.
One sees that the recombination mechanism results in much larger $D$ meson $v_2$ than fragmentation due to the fact that the recombination process brings the flow of light quarks from the hydrodynamic medium into the formation of heavy flavor hadrons. 
Note that in our result, we do not observe significant increase of $D$ meson $v_2$ when combining fragmentation and recombination mechanisms. This may be due to a combinational effect of the initial parton spectra, the momentum dependence of the Wigner function, and the radial flow developed in the QGP medium.

While our calculation seems to underestimate the data of $D$ meson elliptic flow $v_2$, many uncertainties still exist. For instance, if we adopt the KLN initial condition for the hydrodynamic evolution as shown in Fig. \ref{fig:LHCv2init}, we obtain an increase of $D$ meson flow by 25\% due to a larger eccentricity for the initial density profile. Note that changing initial conditions with a larger eccentricity does not affect the overall suppression of $D$ mesons \cite{Cao:2012jt}. In addition, our present study has only been coupled to the hydrodynamic medium evolved from an event-by-event averaged smooth initial conditions; the use of complete event-by-event hydrodynamic evolution profiles will provide a closer comparison to the realistic experimental observation and may also affect the final elliptic flow $v_2$ for heavy flavor hadrons. Such an effort will be explored in a later study.

\begin{figure}[tb]
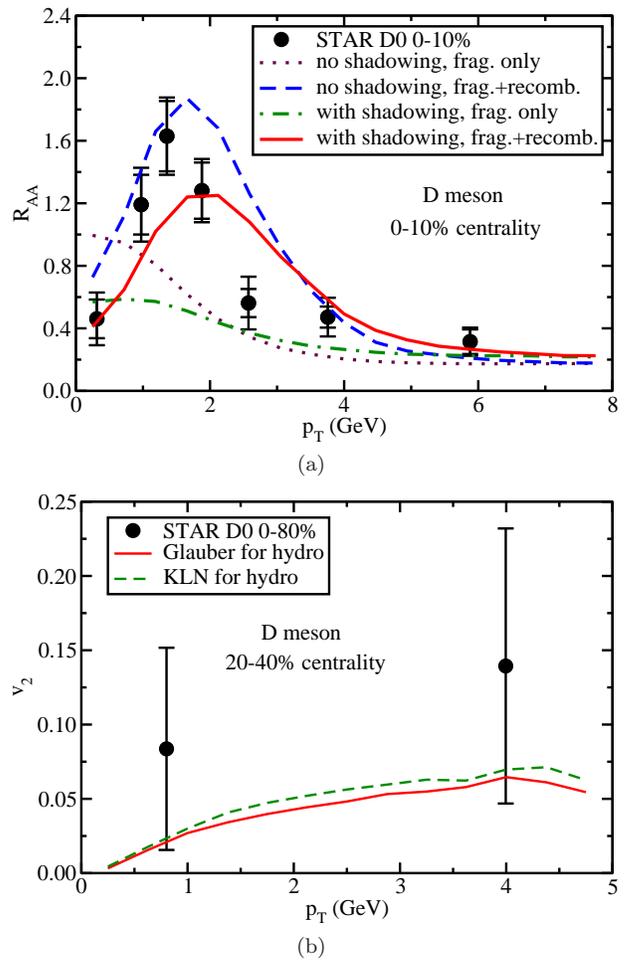

 \subfigure[]{\label{fig:RAARHICD}
      \epsfig{file=RAA_DD6-RHIC-0-10.eps, width=0.45\textwidth, clip=}}
 \subfigure[]{\label{fig:v2RHICD}
      \epsfig{file=RHICDv2.eps, width=0.45\textwidth, clip=}}
 \caption{(Color online) Nuclear modification factor $R_\mathrm{AA}$ (a) and elliptic flow $v_2$ (b) at RHIC.}
 \label{fig:RHICD}
\end{figure}

In Fig. \ref{fig:RHICD} we present the calculation of $D$ meson $R_\mathrm{AA}$ and $v_2$ at RHIC energies in comparison with the data measured by the STAR collaboration. 
We observe that the influence of the nuclear shadowing at RHIC is not as significant as at LHC.
The coalescence mechanism, on the other hand, is found to be more important at RHIC than LHC; one observes the ``bump" structure of the $D$ meson suppression after the incorporation of recombination mechanism in the hadronization process. Our result is consistent with data from the STAR Collaboration.
The results of $D$ meson $v_2$ at RHIC are shown in Fig. \ref{fig:v2RHICD}, where the results for Glauber and KLN hydro initial conditions are compared.
Overall, our model provides a good description of $D$ meson nuclear modification and elliptic flow at RHIC after we take into account the nuclear shadowing effect in the initial heavy quark production, incorporate gluon radiation and elastic collisions for heavy quark evolution and energy loss in medium, and utilize a hybrid model of fragmentation and recombination for heavy quark hadronization process.

\begin{figure}[tb]
 \subfigure[]{\label{fig:RAALHCB}
      \epsfig{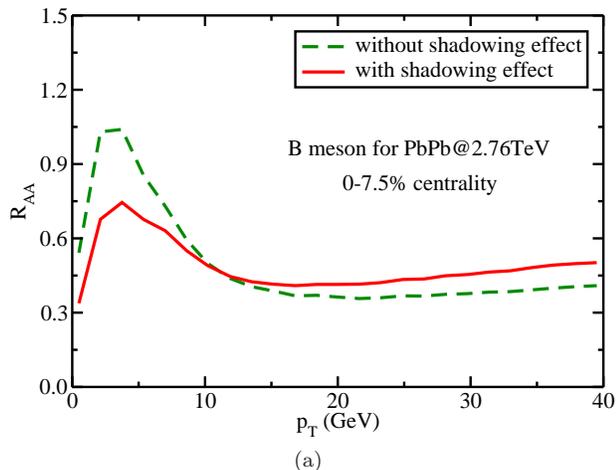}}
 \subfigure[]{\label{fig:v2LHCB}
      \epsfig{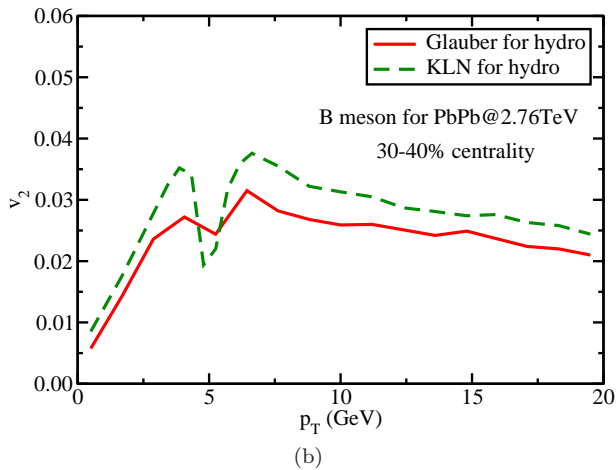}}
 \caption{(Color online) Nuclear modification factor $R_\mathrm{AA}$ (a) and elliptic flow $v_2$ (b) at the LHC. }
 \label{fig:LHCB}
\end{figure}

\begin{figure}[tb]
 \subfigure[]{\label{fig:RAARHICB}
      \epsfig{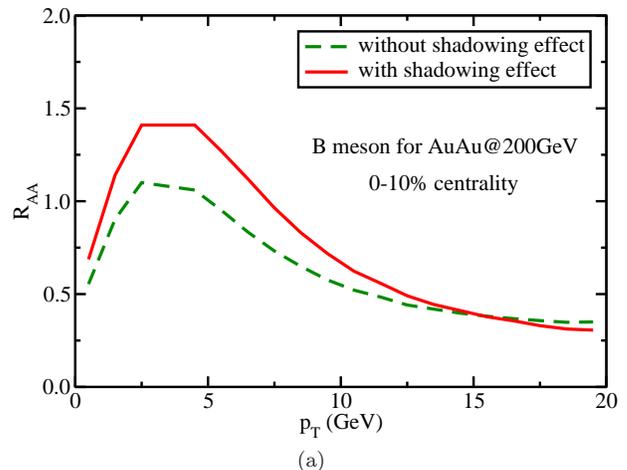}}
 \subfigure[]{\label{fig:v2RHICB}
      \epsfig{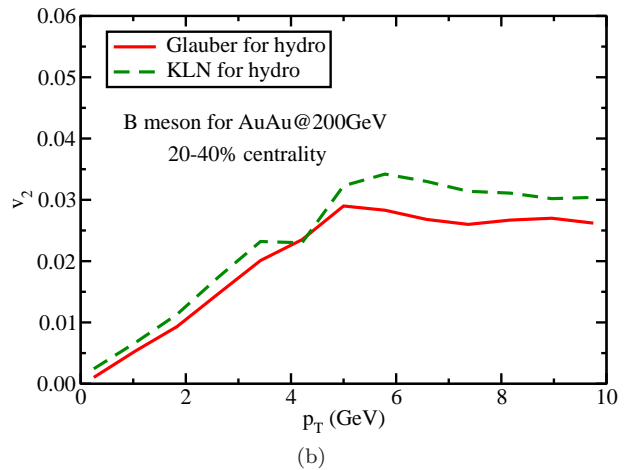}}
 \caption{(Color online) Nuclear modification factor $R_\mathrm{AA}$ (a) and elliptic flow $v_2$ (b) at RHIC. }
 \label{fig:RHICB}
\end{figure}

In Fig. \ref{fig:LHCB} and Fig. \ref{fig:RHICB}, we provide the predictions of the nuclear modification factor and elliptic flow for $B$ mesons at the LHC and RHIC energies. In these two plots, we have included both fragmentation and coalescence mechanisms for bottom quark hadronization. The effects of the nuclear shadowing and different hydrodynamic initial conditions on the final $B$ meson $R_\mathrm{AA}$ and $v_2$ are shown for comparison. Due to the larger mass of bottom quark than that of charm quark, the coalescence mechanism plays a more crucial role in its hadronization process. This can be clearly seen in Fig. \ref{fig:Pcoal} and Fig. \ref{fig:rec-frag}: bottom quarks have much larger recombination probability over a wider $p_\mathrm{T}$ range than charm quarks. As a result, we observe the ``bump'' structure of the $B$ meson $R_\mathrm{AA}$ for both LHC and RHIC. The slight ``dips" in the $B$ meson $v_2$ around 5~GeV in Fig. \ref{fig:v2LHCB} and Fig. \ref{fig:v2RHICB} result from the transition from the regime where collisional energy loss dominates the heavy quark motion to the regime where radiative energy loss takes over. For more details about the relative contributions from different energy loss mechanisms to $R_\mathrm{AA}$ and $v_2$, one may refer to our previous calculation \cite{Cao:2012au}.

\section{Summary and outlook}
\label{sec:summary}

In this work, we have studied the energy loss of heavy quarks and the nuclear modification of heavy flavor mesons in relativistic heavy-ion collisions within the framework of a Langevin approach. To incorporate the contribution from the radiative energy loss, an additional force term has been introduced into the Langevin equation to describe the recoil exerted on heavy quarks due to gluon radiation in which the momenta of the radiated gluons are simulated with the higher-twist energy loss calculation. Within this improved model, we have studied the evolution of heavy quarks propagating through the hot and dense nuclear matter produced in high energy nucleus-nucleus collisions at both LHC and RHIC.

To obtain the final heavy meson spectra and study their modification in high energy nucleus-nucleus collisions, we first initialize our heavy quarks with a leading-order pQCD calculation with the inclusion of the nuclear shadowing effect. The space-time evolution profiles of the hot and dense fireball that heavy quarks traverse have been obtained from a viscous (2+1)-dimensional hydrodynamic model that has been tuned to describe the bulk observables. The hadronization process of heavy quarks after traversing the dense medium has been performed via a hybrid fragmentation plus recombination model.

With our calculation we have demonstrated that the medium-induced gluon radiation contributes significantly to heavy quark energy loss, especially at high energies. The nuclear shadowing has been shown to suppress $D$ meson $R_\mathrm{AA}$ at low $p_\mathrm{T}$ and enhance $R_\mathrm{AA}$ at high $p_\mathrm{T}$. The recombination mechanism has been implemented along with fragmentation for the heavy quark hadronization process and we have found that the inclusion of recombination may increase both $R_\mathrm{AA}$ and $v_2$ of $D$ mesons at intermediate $p_\mathrm{T}$. The effect of different choices of hydrodynamics initial conditions on the final $D$ and $B$ meson elliptic flow has also been investigated. Utilizing our improved Langevin approach together with a hybrid model for heavy quark hadronization, we have presented the nuclear modification and elliptic flow of $D$ mesons, which are consistent with the experimental measurements at both LHC and RHIC. Predictions for future $B$ meson measurements have also been provided. Our calculation can be applied to the spectra of non-photonic electron as well. However, considering the sizable systematic uncertainties due to  the relative charm/bottom quark production rates \cite{Armesto:2005mz,Cao:2012jt}, we defer this effort to a future publication.

Our study constitutes an important contribution to the quantitative understanding of heavy quark production, in-medium evolution and hadronization in relativistic heavy-ion collisions. We plan to extend our study in several directions. For instance, we may include both gluon emission and absorption processes simultaneously for a more rigorous treatment of radiative energy loss in the simulation of heavy quark evolution in medium. We will also extend the simulation from event-by-event averaged smooth hydrodynamic initial conditions to the use of the complete event-by-event hydrodynamics. Last but not least, the evolution of heavy flavors before and after the QGP phase are currently approximated by free-streaming. However, the anomalous transport in the pre-equilibrium state \cite{Mrowczynski:1993qm} and the hadronic interaction between heavy mesons and hadron gas \cite{He:2011yi} may affect the final hadron spectra and elliptic flow. We shall address these aspects in future efforts.

\section*{Acknowledgments}

We are grateful to B. M\"uller and S. Moreland for many helpful discussions. We thank the Ohio State University group (Z. Qiu, C. Shen, H. Song and U. Heinz) for providing the corresponding initialization and hydrodynamic evolution codes, and the Texas A\&M University group (K. C. Han, R. Fries, and C. M. Ko) for discussions on the coalescence model. This work was supported by the U.S. Department of Energy Grant No. DE-FG02-05ER41367.

\bibliographystyle{h-physrev5}
\bibliography{SCrefs}

\end{document}